\documentclass[10pt,english,prl,floatfix,superscriptaddress,twocolumn,showpacs,amsmath,amssymb,reprint]{revtex4-1}
\usepackage[T1]{fontenc}
\usepackage[latin9]{inputenc}
\usepackage{color}
\usepackage{amsmath}
\usepackage{graphicx}

\makeatletter
\PassOptionsToPackage{caption=false}{subfig} 
\usepackage{hyperref}
\hypersetup{
breaklinks=true,
colorlinks=true,
citecolor=blue,
linkcolor=blue,
filecolor=blue,
urlcolor=blue
}
\IfFileExists{lmodern.sty}{\usepackage{lmodern}}{}
\makeatother

\usepackage{babel}
\begin{document}

\title{The collisional resonance function in discrete-resonance quasilinear
plasma systems}

\author{V. N. Duarte}
\email{vduarte@pppl.gov}

\address{Princeton Plasma Physics Laboratory, Princeton University, Princeton,
NJ, 08543, USA}

\author{N. N. Gorelenkov}

\address{Princeton Plasma Physics Laboratory, Princeton University, Princeton,
NJ, 08543, USA}

\author{R. B. White}

\address{Princeton Plasma Physics Laboratory, Princeton University, Princeton,
NJ, 08543, USA}

\author{H. L. Berk}

\address{Institute for Fusion Studies, University of Texas, Austin, TX, 78712,
USA}

\date{\today}
\begin{abstract}
A method is developed to analytically determine the resonance broadening
function in quasilinear theory, due to either Krook or Fokker-Planck
scattering collisions of marginally unstable plasma systems where
discrete resonance instabilities are excited without any mode overlap.
It is demonstrated that a quasilinear system that employs the calculated
broadening functions reported here systematically recovers the nonlinear
growth rate and mode saturation levels for near-threshold plasmas
previously calculated from kinetic theory. The distribution function
is also calculated, which enables precise determination of the characteristic
collisional resonance width.
\end{abstract}
\maketitle
The collisional broadening of resonance lines is a universal phenomenon
in physics. For example, in atomic physics, collisions lead to abrupt
changes in phase and plane of vibration, thereby destroying phase
coherence and leading to uncertainty in the associated photon energy.
This leads to broadening of the atoms emission/absorption profile
\citep{Lorentz1906,weisskopf1933breite}. In plasma physics, decoherence
of the orbital motion of resonant particles allows the reduction of
reversible equations of motion into a diffusive system of equations
that governs the resonant particle dynamics without detailed tracking
of the ballistic motion - as is the case in the widely used quasilinear
(QL) formulations of \citep{VedenovSagdeev1961,Drummond_Pines_1962,KaufmanQLPoF1972}.
In spite of being an essential element of the structure of QL theory,
the determination of the appropriate collisional broadening resonance
function has not yet been formulated. In this Letter, we show how
to calculate the collisional resonance function from first principles
and show that its use implies that a QL plasma system automatically
replicates the nonlinear growth rate and the wave saturation levels
calculated from full kinetic theory near marginality \citep{BerkPRL1996,BreizmanPoP1997}.

We shall show how the results of previous works that focused on the
dynamics of plasma systems just above the marginal state for instability
\citep{BerkPRL1996,BreizmanPoP1997} can be interpreted within the
context of QL theory. Ref. \citep{BerkPRL1996} developed a method
that calculated the transition from the linearly unstable regime to
the nonlinear stabilized regime. In their investigation, a cubic nonlinear
time delay equation was derived and applied to a wide variety of plasma
systems (e.g., the bump-on-tail problem in Q-machine-like devices,
alpha particle induced instability that is crucial in burning plasmas
\citep{FasoliPRL1998} and prediction of the emergence of wave frequency
chirping in tokamaks \citep{DuarteAxivPRL}). These studies showed
that, with stochastic mechanisms present, such as collisions and background
turbulence, quasi-steady solutions could be found. Based on these
results, a heuristic QL method was developed  \citep{Berk1995LBQ}
that replicated the results of  these stationary solutions, both near
and far from marginal stability. This model was an extension of the
collisionless QL theory developed by Kaufman \citep{KaufmanQLPoF1972}.
Berk \citep{Berk1995LBQ} suggested intuitive rules, relying on an
arbitrarily chosen shape, for creating a resonance function (i.e.,
an envelope function that weights the strength of the resonant interaction)
that broadened the singular delta functions that appear in Kaufman's
theory. The aim of the present work is to show that just above the
marginal instability state, a systematic QL theory can be developed,
where one obtains a resonance function that integrates to unity, as
physically expected. Without further assumption, what then emerges
is the shape of the resonance function and the mode saturation level,
which replicates the results of the original kinetic calculations
\citep{BerkPRL1996,BreizmanPoP1997}. The predicted saturation level
of the kinetic theory resulted from the derivation of a rather complex
time-delayed integro-differential equation, which turned out to be
identical to the evolution equation previously derived for a shear flow
fluid problem involving Rossby waves \citep{Hickernell1984}. In contrast,
the predictions of this new QL theory is derived from a simple set
of equations which yields a clear understanding of the physical processes
that are taking place. The QL theory that is developed is applicable
to complex, multi-dimensional systems. In particular the new theory
is being applied to whole-device modeling of multiple Alfv\'{e}nic instabilities
that are driven by energetic beams and fusion products in tokamaks
\citep{GorelenkovPoP2019}.

Resonant particles are described via a distribution function $f\left(\varphi,\Omega;t\right)$,
where $\varphi$ is a canonical angle, $\Omega$ is a frequency-like
variable which is a function of the relevant action $J$ (canonically
conjugated to $\varphi$) \citep{BerkPPR1997}, and $t$ is time.
$\Omega=0$ determines the resonance condition. The kinetic equation
for a single resonance is (the generalization of the method for treating
multiple non-overlapping resonances is straightforward, and will be
presented in a subsequent more expansive publication rather than in
this Letter)
\begin{equation}
\frac{\partial f}{\partial t}+\Omega\frac{\partial f}{\partial\varphi}+Re\left(\omega_{b}^{2}e^{i\varphi}\right)\frac{\partial f}{\partial\Omega}=C\left[f,F_{0}\right],
\end{equation}
where the form for the collisional operator $C[f,F_{0}]$ is taken
as either $\nu_{K}\left(F_{0}-f\right)$, which are the creation and
annihilation terms of the Krook model \citep{Krook1954} or $\nu_{scatt}^{3}\partial^{2}\left(f-F_{0}\right)/\partial\Omega^{2}$,
which is the diffusive scattering operator \citep{trubnikov1965particle},
and $\nu_{K}$ and $\nu_{scatt}$ are the effective collision frequencies.
$\omega_{b}$ is the nonlinear trapping (bounce) frequency at a given
resonance, which is proportional to the square root of the mode amplitude.
$F_{0}$ is the distribution function in the absence of wave perturbations.
The distribution can be assumed of the form $f\left(\varphi,\Omega,t\right)=F_{0}\left(\Omega\right)+f_{0}\left(\Omega,t\right)+\underset{n=1}{\overset{\infty}{\sum}}\left(f_{n}\left(\Omega,t\right)e^{in\varphi}+c.c.\right)$
with the ordering $\left|F_{0}'\right|\gg\left|f_{1}'^{(1)}\right|\gg\left|f_{0}'^{(2)}\right|,\left|f_{2}'^{(2)}\right|$
\citep{SanzNF2018}. The prime denotes the derivative with respect
to $\Omega$ while the superscript denotes the order in the wave amplitude
(equivalently, in orders of $\omega_{b}^{2}$). Then the $f_{n}$
satisfy

\begin{equation}
\begin{array}{c}
\frac{\partial f_{n}}{\partial t}+in\Omega f_{n}+\frac{1}{2}\left(\omega_{b}^{2}f'_{n-1}+\omega_{b}^{2*}f'_{n+1}\right)=\\
=\left\{ -\nu_{K}f_{n},\nu_{scatt}^{3}f_{n}''\right\} 
\end{array}\label{eq:KineticEqForN}
\end{equation}
where the brackets on the right hand side denote either Krook or scattering
operators. Sufficiently close to the linear instability threshold,
with even moderate collisionality, \textcolor{black}{$\nu_{K,scatt}/\left(\gamma_{L,0}-\gamma_{d}\right)\gg1$
is satisfied ($\gamma_{L,0}$ is the mode linear growth rate at $t=0$
and $\gamma_{d}$ is the background damping rate). In this case, the
detailed time history is not essential for the description of the
system's dynamics \citep{duarte2018analytical}. Then, to lowest order
in $\omega_{b}^{2}/\nu_{K,scatt}^{2}$ one can disregard the time
derivative in} \eqref{eq:KineticEqForN}. Therefore, the principal
time dependency contribution to $f_{n}$ comes from $\omega_{b}(t)$
rather than from a delayed time integral over the particle distribution\textquoteright s
time history.

Starting with the Krook case, to first order in \textcolor{black}{$\omega_{b}^{2}/\nu_{K}^{2}$},
Eq. \eqref{eq:KineticEqForN} gives 
\begin{equation}
f_{1}=\frac{\omega_{b}^{2}F_{0}'}{2\left(i\Omega+\nu_{K}\right)}.\label{eq:f1K}
\end{equation}
Noting that the reality constraint implies $f_{-1}=f_{1}^{*}$, to
second order in \textcolor{black}{$\omega_{b}^{2}/\nu_{K}^{2}$},
\eqref{eq:KineticEqForN} gives 
\begin{equation}
\frac{\partial f_{0}}{\partial t}+\frac{1}{2}\left(\omega_{b}^{2}\left[f'_{1}\right]^{*}+\omega_{b}^{2*}f'_{1}\right)=-\nu_{K}f_{0}.\label{eq:f0K}
\end{equation}
Defining the angle-independent distribution as \textcolor{black}{$f\left(\Omega,t\right)\equiv F_{0}\left(\Omega\right)+f_{0}\left(\Omega,t\right)$
and }noting that by construction $\partial F_{0}/\partial t=0$ and
$\left|F_{0}'\right|\gg\left|f_{0}'\right|$, \textcolor{black}{one
then obtains from Eqs. \eqref{eq:f1K} and }\eqref{eq:f0K}\textcolor{black}{{}
that the relaxation of $f\left(\Omega,t\right)$} is governed by the
diffusion equation
\begin{equation}
\frac{\partial f\left(\Omega,t\right)}{\partial t}-\frac{\pi}{2}\frac{\partial}{\partial\Omega}\left[\left|\omega_{b}^{2}\right|^{2}\mathcal{R}\left(\Omega\right)\frac{\partial f\left(\Omega,t\right)}{\partial\Omega}\right]=C\left[f,F_{0}\right]\label{eq:QLEq}
\end{equation}
where, for the Krook case, $\mathcal{R}\left(\Omega\right)$ is

\begin{equation}
\mathcal{R}_{K}(\Omega)=\frac{1}{\pi\nu_{K}\left(1+\Omega^{2}/\nu_{K}^{2}\right)}.\label{eq:WKrook}
\end{equation}
A somewhat similar procedure can be employed for the scattering case.
To first order in \textcolor{black}{$\omega_{b}^{2}/\nu_{scatt}^{2}$},
we integrate Eq. \eqref{eq:KineticEqForN} along the characteristics,
which gives 
\begin{equation}
f_{1}=\frac{iF'_{0}\omega_{b}^{2}\left(t\right)}{2\nu_{scatt}}\int_{-\infty}^{0}dse^{i\frac{\Omega}{\nu_{scatt}}s}e^{s^{3}/3}.\label{eq:f1scatt}
\end{equation}
Eq. \eqref{eq:f1scatt} is then iterated in \eqref{eq:KineticEqForN}
to second order in \textcolor{black}{$\omega_{b}^{2}/\nu_{scatt}^{2}$}.
Again, using that $\partial F_{0}/\partial t=0$ and $\left|F_{0}'\right|\gg\left|f_{0}'\right|$,
it is readily found that \textcolor{black}{$f\left(\Omega,t\right)\equiv F_{0}\left(\Omega\right)+f_{0}\left(\Omega,t\right)$
for the scattering case also satisfies an equation of the form of
Eq. \eqref{eq:QLEq}, with}

\begin{figure}
\begin{centering}
\includegraphics[scale=0.32]{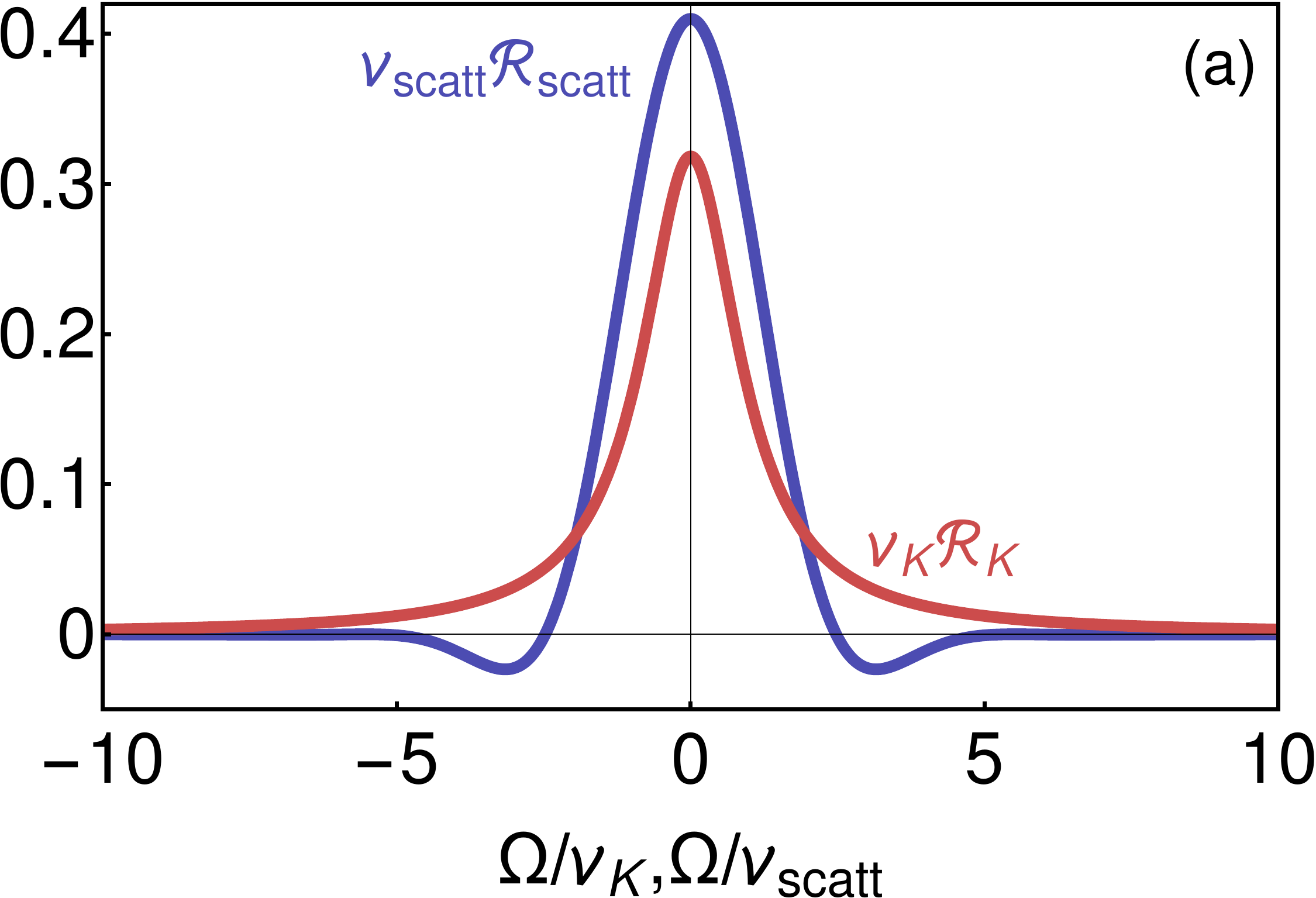}
\par\end{centering}
\centering{}\includegraphics[scale=0.32]{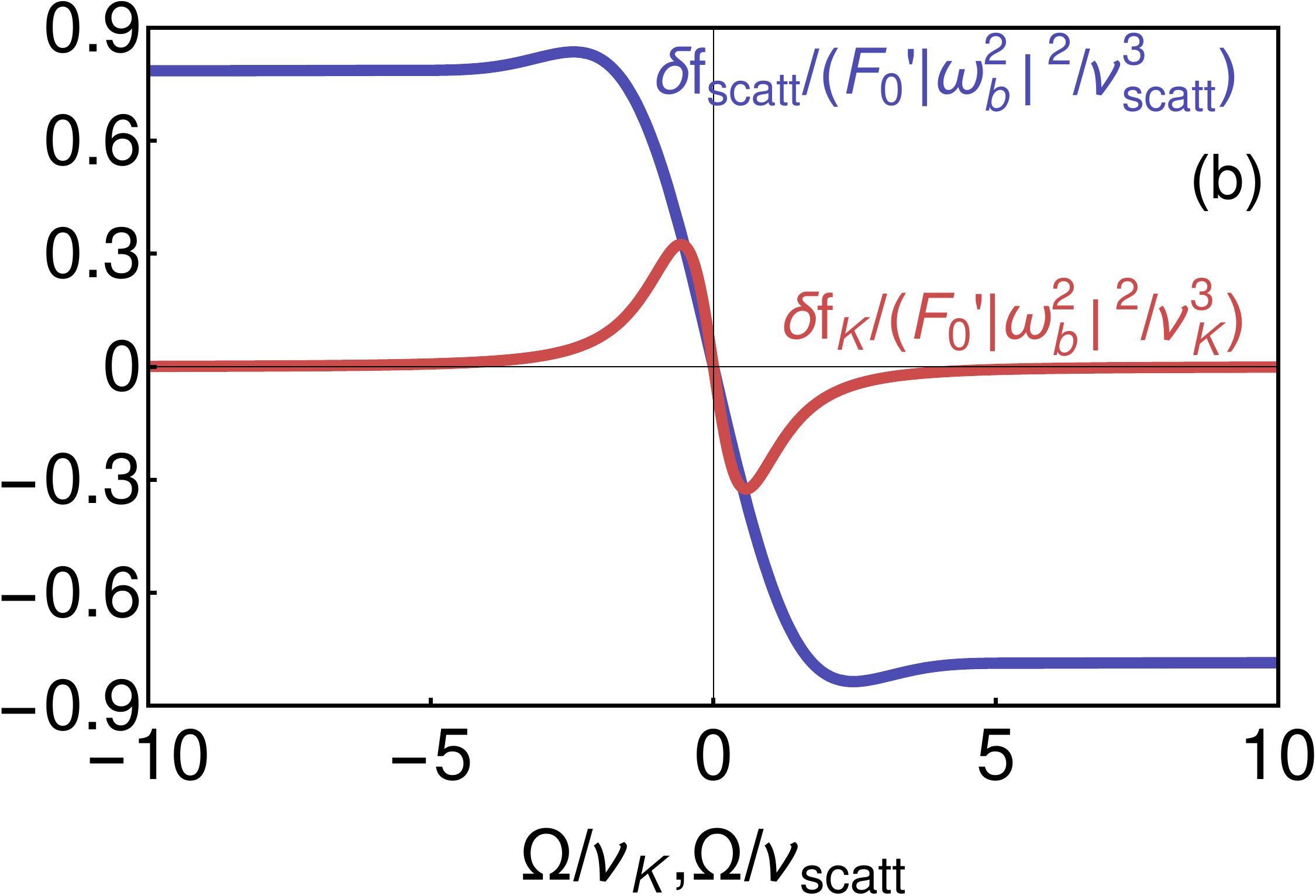}\caption{(a) Resonance function (Eqs. \ref{eq:WKrook} and \textcolor{black}{\eqref{eq:Wscatt}})
and (b) $\delta f=f-F_{0}$ (Eqs. \eqref{eq:deltafK} and \eqref{eq:deltafscatt})
vs. normalized frequency variable. The red and blue curves correspond
to the Krook and scattering cases, respectively. The full width at
half maximum of the resonance function in part (a) is $\Delta\Omega=2\nu_{K}$
for Krook and $\Delta\Omega\protect\cong2.58\nu_{scatt}$ for the
scattering case. The separation between the two peaks of each curve
for $\delta f$ in plot (b)  is $\Delta\Omega=2\nu_{K}/\sqrt{3}$
for Krook and $\Delta\Omega\protect\cong4.95\nu_{scatt}$ for the
scattering case.\label{FigWindFunc}}
\end{figure}

\textcolor{black}{
\begin{equation}
\mathcal{R}_{scatt}\left(\Omega\right)=\frac{1}{\pi\nu_{scatt}}\int_{0}^{\infty}ds\:\cos\left(\frac{\Omega s}{\nu_{scatt}}\right)e^{-s^{3}/3}.\label{eq:Wscatt}
\end{equation}
}The resonance functions \eqref{eq:WKrook} and \textcolor{black}{\eqref{eq:Wscatt}
are} plotted in Fig. \ref{FigWindFunc}(a). The property $\int_{-\infty}^{\infty}\mathcal{F}(\Omega)d\Omega=1$,
expected for functions that replace a delta function, is automatically
satisfied by both forms of the resonance function. For a self-consistent
description, the QL diffusion Eq. \textcolor{black}{\eqref{eq:QLEq}
must be solved simultaneously with the Eq. for amplitude evolution,
$d\left|\omega_{b}^{2}\right|^{2}/dt=2\left(\gamma_{L}\left(t\right)-\gamma_{d}\right)\left|\omega_{b}^{2}\right|^{2}$,
and for the growth rate, $\gamma_{L}\left(t\right)=\frac{\pi}{4}\int_{-\infty}^{\infty}d\Omega\mathcal{R}\frac{\partial f\left(\Omega,t\right)}{\partial\Omega}$.}

\textcolor{black}{Interestingly, functions similar to }\eqref{eq:WKrook}
and \textcolor{black}{\eqref{eq:Wscatt} appear in the context of
broadening of atomic emission lines }- their equivalent \textcolor{black}{are
Eq. 12 of \citep{vanVleck1945shape} and Eq. 5.68 (with $p=1$) of
\citep{peach1981theory}, respectively. Eq. \eqref{eq:Wscatt} has
the} same form of the function calculated by Dupree \textcolor{black}{\citep{Dupree1966}
in a different context, namely} in the study of strong turbulence
theory, where a dense spectrum of fluctuations diffuse particles away
from their free-streaming trajectories (see Ref. \textcolor{black}{\citep{krommes2002fundamental}}
for a review covering broadening theories in strong turbulence). In
that case, a renormalized average propagator was introduced and the
cubic term in the argument of the exponential is proportional to a
collisionless diffusion coefficient.

A concern might arise about the physical significance of a resonance
function that is negative in a part of its domain, as is shown in
Fig. \ref{FigWindFunc}(a)\textcolor{black}{{} for the function \eqref{eq:Wscatt}}.
\textcolor{black}{We note that for the problem treated in the present
work, the collisional diffusion ensures that the overall diffusion
coefficient in Eq. \eqref{eq:QLEq} is always positive. In Dupree's
case, the assumed overlapping turbulent dense spectrum ensures positivity
over the entire phase-space domain. }

\textcolor{black}{To leading order near marginal instability, there
emerges the following higher order steady state distribution functions
($\delta f\equiv f\left(\Omega,t\right)-F_{0}\left(\Omega\right)$)
from Eq. \eqref{eq:QLEq}. For the Krook model, it has the form}

\textcolor{black}{
\begin{equation}
\delta f_{K}=-\frac{\left|\omega_{b}^{2}\right|^{2}}{\nu_{K}^{3}}\frac{\partial F_{0}}{\partial\Omega}\frac{\Omega/\nu_{K}}{\left(1+\Omega^{2}/\nu_{K}^{2}\right)^{2}}\label{eq:deltafK}
\end{equation}
while for the diffusive scattering model,
\begin{equation}
\delta f_{scatt}=-\frac{\left|\omega_{b}^{2}\right|^{2}}{2\nu_{scatt}^{3}}\frac{\partial F_{0}}{\partial\Omega}\int_{0}^{\infty}\frac{ds}{s}\sin\left(\frac{\Omega s}{\nu_{scatt}}\right)e^{-s^{3}/3}\label{eq:deltafscatt}
\end{equation}
}

Fig. \ref{FigWindFunc}(b) shows the forms for the marginally unstable
$\delta f$. These forms can be useful for code verification akin
to studies reported in Ref. \citep{WhiteDuarteGorMengPoP2019}.  Fig.
\ref{FigWindFunc}(b) is valid in the vicinity of the resonance -
its behavior far from the resonance would then be determined by the
boundary conditions one imposes to \textcolor{black}{Eq. \eqref{eq:QLEq}
}.

\textcolor{black}{We now demonstrate that near the instability threshold,
the QL theory together with the calculated resonance functions (}\eqref{eq:WKrook}\textcolor{black}{{}
and \eqref{eq:Wscatt}) replicates the same saturation levels calculated
by nonlinear theory }\citep{BerkPRL1996,BreizmanPoP1997}\textcolor{black}{.}
\textcolor{black}{Let us start with Eq. \eqref{eq:QLEq} for the Krook
case. To leading order, it can be written as $-\frac{\pi}{2}\left|\omega_{b}^{2}\right|^{2}\frac{\partial F_{0}}{\partial\Omega}\frac{\partial\mathcal{R}}{\partial\Omega}=\nu_{K}\left(F_{0}-f\right)$,
since the marginality condition implies $\omega_{b}\ll\nu_{K},\nu_{scatt}$.
Differentiating with respect to $\Omega$, then multiplying by $\mathcal{R}$
and integrating over $\Omega$, we get:
\begin{equation}
\left|\omega_{b}^{2}\right|^{2}\int_{-\infty}^{\infty}\mathcal{R}\frac{\partial^{2}\mathcal{R}}{\partial\Omega^{2}}d\Omega=-\frac{2\nu_{K}}{\pi\frac{\partial F_{0}}{\partial\Omega}}\int_{-\infty}^{\infty}\mathcal{R}\left(\frac{\partial F_{0}}{\partial\Omega}-\frac{\partial f}{\partial\Omega}\right)d\Omega\label{eq:ProofKrook}
\end{equation}
Note that, because $\mathcal{R}$ vanishes at $\pm\infty$, integration
by parts of the left hand side leads to $\int_{-\infty}^{\infty}\mathcal{R}\frac{\partial^{2}\mathcal{R}}{\partial\Omega^{2}}d\Omega=-\int_{-\infty}^{\infty}\left(\frac{\partial\mathcal{R}}{\partial\Omega}\right)^{2}d\Omega=\frac{-1}{4\pi\nu_{K}^{3}}$
(the last equality follows from using the function given in Eq. \eqref{eq:WKrook}).
Noting that the initial growth rate (at $t=0$) is defined as $\gamma_{L,0}=\frac{\pi}{4}\int_{-\infty}^{\infty}d\Omega\mathcal{R}\frac{\partial F_{0}}{\partial\Omega}$
and the dynamical QL growth rate is $\gamma_{L}\left(t\right)=\frac{\pi}{4}\int_{-\infty}^{\infty}d\Omega\mathcal{R}\frac{\partial f\left(\Omega,t\right)}{\partial\Omega}$,
it follows from Eq. \eqref{eq:ProofKrook} that $\gamma_{L}\left(t\right)=\gamma_{L,0}\left(1-\left|\omega_{b}^{2}(t)\right|^{2}/8\nu_{K}^{4}\right)$.
At saturation, i.e., when $\gamma_{L}=\gamma_{d}$, then $\left|\omega_{b,sat}\right|=8^{1/4}\left(1-\gamma_{d}/\gamma_{L,0}\right)^{1/4}\nu_{K}$,
which is the same saturation level as the one predicted by the kinetic
time-delayed integral nonlinear equation }\citep{BerkPRL1996}.

\textcolor{black}{A slightly different procedure can be employed for
the scattering case, for which the QL diffusion Eq. \eqref{eq:QLEq}
can be written to leading order as $-\frac{\pi}{2}\left|\omega_{b}^{2}\right|^{2}\frac{\partial F_{0}}{\partial\Omega}\frac{\partial\mathcal{R}}{\partial\Omega}=\nu_{scatt}^{3}\frac{\partial^{2}\left(f-F_{0}\right)}{\partial\Omega^{2}}$.
Integrating over $\Omega$, multiplying both sides by $\mathcal{R}$
and integrating over $\Omega$, one obtains
\begin{equation}
\left|\omega_{b}^{2}\right|^{2}\int_{-\infty}^{\infty}\mathcal{R}^{2}d\Omega=\frac{2\nu_{scatt}^{3}}{\pi\frac{\partial F_{0}}{\partial\Omega}}\int_{-\infty}^{\infty}\mathcal{R}\left(\frac{\partial F_{0}}{\partial\Omega}-\frac{\partial f}{\partial\Omega}\right)d\Omega\label{eq:ProofScatt}
\end{equation}
The integration on the left hand side can be analytically performed
using Eq. \eqref{eq:Wscatt}, which gives $\int_{-\infty}^{\infty}\mathcal{R}^{2}d\Omega=\frac{2}{\pi\nu_{scatt}}\left[\Gamma\left(\frac{1}{3}\right)\left(\frac{3}{2}\right)^{1/3}\frac{1}{6}\right]^{-4}$.
Using the definitions $\gamma_{L,0}=\frac{\pi}{4}\int_{-\infty}^{\infty}d\Omega\mathcal{R}\frac{\partial F_{0}}{\partial\Omega}$
and $\gamma_{L}\left(t\right)=\frac{\pi}{4}\int_{-\infty}^{\infty}d\Omega\mathcal{R}\frac{\partial f\left(\Omega,t\right)}{\partial\Omega}$,
then one obtains from Eq. \eqref{eq:ProofScatt} that $\gamma_{L}\left(t\right)=\gamma_{L,0}\left[1-\left|\omega_{b}^{2}(t)\right|^{2}\Gamma\left(1/3\right)\left(3/2\right)^{1/3}/\left(6\nu_{scatt}^{4}\right)\right]$.
At saturation, when $\gamma_{L}=\gamma_{d}$, then $\left|\omega_{b,sat}\right|\simeq1.18\left(1-\gamma_{d}/\gamma_{L,0}\right)^{1/4}\nu_{scatt}$,
which is the same as what follows from nonlinear kinetic theory }\citep{BreizmanPoP1997}\textcolor{black}{{}
QED. }

\textcolor{black}{The limit $\nu_{K,scatt}/\left(\gamma_{L,0}-\gamma_{d}\right)\gg1$,
when the detailed time history becomes unimportant, allows for the
derivation of the analytical expression for the nonlinear growth rate
}$\gamma_{NL}\left(t\right)=\gamma_{L,0}\left(1-\alpha\left|\omega_{b}^{2}(t)\right|^{2}\right)$
\textcolor{black}{\citep{duarte2018analytical}}, where $\alpha=\left(8\nu_{K}^{4}\right)^{-1}$
for the Krook case and $\alpha=\Gamma\left(1/3\right)\left(3/2\right)^{1/3}/\left(6\nu_{scatt}^{4}\right)$
for the scattering case. \textcolor{black}{Comparison with the above
expressions for the calculated QL growth rates imply that they are
equal to the nonlinear growth rate at all times for both collisional
cases.}

\textcolor{black}{In conclusion, it has been demonstrated that near
marginal stability, the systematic QL transport theory we developed
replicates the identical growth rates and saturation levels as predicted
by }a significantly more complex\textcolor{black}{{} nonlinear kinetic
theory }based on solving a time delayed integro-differential equation\textcolor{black}{.
The demonstration did not rely on any assumption for the specific
form of the distribution.} We note that our demonstration assumed
that the overall system is governed by a QL equation that self-consistently
embodies collisional effects via a resonance function that was previously
determined from first principles \textcolor{black}{(}\eqref{eq:WKrook}\textcolor{black}{{}
and \eqref{eq:Wscatt})}. However, a QL theory, being a reduced framework,
does not contain all the relevant information as to the detailed angle-resolved
distribution function. Hence, in work to be shown elsewhere, we have
also developed an alternative formal approach, that produces additional
structure as part of the perturbed distribution function that is not
described by the coarse-grained QL theory. However, we have shown
that this additional structure does not alter the nonlinear corrections
to the field amplitude, predicted by the QL theory we report here.
A description of the results of this more general approach will be
given in a later more detailed paper.

This work was supported by the US Department of Energy under contract
DE-AC02-09CH11466.

\bibliographystyle{apsrev4-1}

\begin{thebibliography}{22}%
\makeatletter
\providecommand \@ifxundefined [1]{%
 \@ifx{#1\undefined}
}%
\providecommand \@ifnum [1]{%
 \ifnum #1\expandafter \@firstoftwo
 \else \expandafter \@secondoftwo
 \fi
}%
\providecommand \@ifx [1]{%
 \ifx #1\expandafter \@firstoftwo
 \else \expandafter \@secondoftwo
 \fi
}%
\providecommand \natexlab [1]{#1}%
\providecommand \enquote  [1]{``#1''}%
\providecommand \bibnamefont  [1]{#1}%
\providecommand \bibfnamefont [1]{#1}%
\providecommand \citenamefont [1]{#1}%
\providecommand \href@noop [0]{\@secondoftwo}%
\providecommand \href [0]{\begingroup \@sanitize@url \@href}%
\providecommand \@href[1]{\@@startlink{#1}\@@href}%
\providecommand \@@href[1]{\endgroup#1\@@endlink}%
\providecommand \@sanitize@url [0]{\catcode `\\12\catcode `\$12\catcode
  `\&12\catcode `\#12\catcode `\^12\catcode `\_12\catcode `\%12\relax}%
\providecommand \@@startlink[1]{}%
\providecommand \@@endlink[0]{}%
\providecommand \url  [0]{\begingroup\@sanitize@url \@url }%
\providecommand \@url [1]{\endgroup\@href {#1}{\urlprefix }}%
\providecommand \urlprefix  [0]{URL }%
\providecommand \Eprint [0]{\href }%
\providecommand \doibase [0]{http://dx.doi.org/}%
\providecommand \selectlanguage [0]{\@gobble}%
\providecommand \bibinfo  [0]{\@secondoftwo}%
\providecommand \bibfield  [0]{\@secondoftwo}%
\providecommand \translation [1]{[#1]}%
\providecommand \BibitemOpen [0]{}%
\providecommand \bibitemStop [0]{}%
\providecommand \bibitemNoStop [0]{.\EOS\space}%
\providecommand \EOS [0]{\spacefactor3000\relax}%
\providecommand \BibitemShut  [1]{\csname bibitem#1\endcsname}%
\let\auto@bib@innerbib\@empty
%</preamble>
\bibitem [{\citenamefont {Lorentz}(1906)}]{Lorentz1906}%
  \BibitemOpen
  \bibfield  {author} {\bibinfo {author} {\bibfnamefont {H.~A.}\ \bibnamefont
  {Lorentz}},\ }\href@noop {} {\bibfield  {journal} {\bibinfo  {journal} {Proc.
  Amst. Acad. Sci.}\ }\textbf {\bibinfo {volume} {8}},\ \bibinfo {pages} {591}
  (\bibinfo {year} {1906})}\BibitemShut {NoStop}%
\bibitem [{\citenamefont {Weisskopf}(1933)}]{weisskopf1933breite}%
  \BibitemOpen
  \bibfield  {author} {\bibinfo {author} {\bibfnamefont {V.}~\bibnamefont
  {Weisskopf}},\ }\href@noop {} {\bibfield  {journal} {\bibinfo  {journal}
  {Phys. Zeit.}\ }\textbf {\bibinfo {volume} {34}},\ \bibinfo {pages} {1}
  (\bibinfo {year} {1933})}\BibitemShut {NoStop}%
\bibitem [{\citenamefont {Vedenov}\ \emph {et~al.}(1961)\citenamefont
  {Vedenov}, \citenamefont {Velikhov},\ and\ \citenamefont
  {Sagdeev}}]{VedenovSagdeev1961}%
  \BibitemOpen
  \bibfield  {author} {\bibinfo {author} {\bibfnamefont {A.~A.}\ \bibnamefont
  {Vedenov}}, \bibinfo {author} {\bibfnamefont {E.~P.}\ \bibnamefont
  {Velikhov}}, \ and\ \bibinfo {author} {\bibfnamefont {R.~Z.}\ \bibnamefont
  {Sagdeev}},\ }\href {http://stacks.iop.org/0038-5670/4/i=2/a=A12} {\bibfield
  {journal} {\bibinfo  {journal} {Sov. Phys. Uspekhi}\ }\textbf {\bibinfo
  {volume} {4}},\ \bibinfo {pages} {332} (\bibinfo {year} {1961})}\BibitemShut
  {NoStop}%
\bibitem [{\citenamefont {Drummond}\ and\ \citenamefont
  {Pines}(1962)}]{Drummond_Pines_1962}%
  \BibitemOpen
  \bibfield  {author} {\bibinfo {author} {\bibfnamefont {W.}~\bibnamefont
  {Drummond}}\ and\ \bibinfo {author} {\bibfnamefont {D.}~\bibnamefont
  {Pines}},\ }\href@noop {} {\bibfield  {journal} {\bibinfo  {journal} {Nucl.
  Fusion}\ }\textbf {\bibinfo {volume} {Suppl. 2, Pt. 3}} (\bibinfo {year}
  {1962})}\BibitemShut {NoStop}%
\bibitem [{\citenamefont {Kaufman}(1972)}]{KaufmanQLPoF1972}%
  \BibitemOpen
  \bibfield  {author} {\bibinfo {author} {\bibfnamefont {A.~N.}\ \bibnamefont
  {Kaufman}},\ }\href {\doibase http://dx.doi.org/10.1063/1.1694031} {\bibfield
   {journal} {\bibinfo  {journal} {Phys. Fluids}\ }\textbf {\bibinfo {volume}
  {15}},\ \bibinfo {eid} {1063} (\bibinfo {year} {1972})}\BibitemShut {NoStop}%
\bibitem [{\citenamefont {Berk}\ \emph {et~al.}(1996)\citenamefont {Berk},
  \citenamefont {Breizman},\ and\ \citenamefont {Pekker}}]{BerkPRL1996}%
  \BibitemOpen
  \bibfield  {author} {\bibinfo {author} {\bibfnamefont {H.~L.}\ \bibnamefont
  {Berk}}, \bibinfo {author} {\bibfnamefont {B.~N.}\ \bibnamefont {Breizman}},
  \ and\ \bibinfo {author} {\bibfnamefont {M.}~\bibnamefont {Pekker}},\ }\href
  {\doibase 10.1103/PhysRevLett.76.1256} {\bibfield  {journal} {\bibinfo
  {journal} {Phys. Rev. Lett.}\ }\textbf {\bibinfo {volume} {76}},\ \bibinfo
  {pages} {1256} (\bibinfo {year} {1996})}\BibitemShut {NoStop}%
\bibitem [{\citenamefont {Breizman}\ \emph {et~al.}(1997)\citenamefont
  {Breizman}, \citenamefont {Berk}, \citenamefont {Pekker}, \citenamefont
  {Porcelli}, \citenamefont {Stupakov},\ and\ \citenamefont
  {Wong}}]{BreizmanPoP1997}%
  \BibitemOpen
  \bibfield  {author} {\bibinfo {author} {\bibfnamefont {B.~N.}\ \bibnamefont
  {Breizman}}, \bibinfo {author} {\bibfnamefont {H.~L.}\ \bibnamefont {Berk}},
  \bibinfo {author} {\bibfnamefont {M.~S.}\ \bibnamefont {Pekker}}, \bibinfo
  {author} {\bibfnamefont {F.}~\bibnamefont {Porcelli}}, \bibinfo {author}
  {\bibfnamefont {G.~V.}\ \bibnamefont {Stupakov}}, \ and\ \bibinfo {author}
  {\bibfnamefont {K.~L.}\ \bibnamefont {Wong}},\ }\href {\doibase
  http://dx.doi.org/10.1063/1.872286} {\bibfield  {journal} {\bibinfo
  {journal} {Phys. Plasmas}\ }\textbf {\bibinfo {volume} {4}},\ \bibinfo
  {pages} {1559} (\bibinfo {year} {1997})}\BibitemShut {NoStop}%
\bibitem [{\citenamefont {Fasoli}\ \emph {et~al.}(1998)\citenamefont {Fasoli},
  \citenamefont {Breizman}, \citenamefont {Borba}, \citenamefont {Heeter},
  \citenamefont {Pekker},\ and\ \citenamefont {Sharapov}}]{FasoliPRL1998}%
  \BibitemOpen
  \bibfield  {author} {\bibinfo {author} {\bibfnamefont {A.}~\bibnamefont
  {Fasoli}}, \bibinfo {author} {\bibfnamefont {B.~N.}\ \bibnamefont
  {Breizman}}, \bibinfo {author} {\bibfnamefont {D.}~\bibnamefont {Borba}},
  \bibinfo {author} {\bibfnamefont {R.~F.}\ \bibnamefont {Heeter}}, \bibinfo
  {author} {\bibfnamefont {M.~S.}\ \bibnamefont {Pekker}}, \ and\ \bibinfo
  {author} {\bibfnamefont {S.~E.}\ \bibnamefont {Sharapov}},\ }\href {\doibase
  10.1103/PhysRevLett.81.5564} {\bibfield  {journal} {\bibinfo  {journal}
  {Phys. Rev. Lett.}\ }\textbf {\bibinfo {volume} {81}},\ \bibinfo {pages}
  {5564} (\bibinfo {year} {1998})}\BibitemShut {NoStop}%
\bibitem [{\citenamefont {Duarte}\ \emph {et~al.}(2017)\citenamefont {Duarte},
  \citenamefont {Berk}, \citenamefont {Gorelenkov}, \citenamefont {Heidbrink},
  \citenamefont {Kramer}, \citenamefont {Nazikian}, \citenamefont {Pace},
  \citenamefont {Podest{\`{a}}}, \citenamefont {Tobias},\ and\ \citenamefont
  {Van~Zeeland}}]{DuarteAxivPRL}%
  \BibitemOpen
  \bibfield  {author} {\bibinfo {author} {\bibfnamefont {V.~N.}\ \bibnamefont
  {Duarte}}, \bibinfo {author} {\bibfnamefont {H.~L.}\ \bibnamefont {Berk}},
  \bibinfo {author} {\bibfnamefont {N.~N.}\ \bibnamefont {Gorelenkov}},
  \bibinfo {author} {\bibfnamefont {W.~W.}\ \bibnamefont {Heidbrink}}, \bibinfo
  {author} {\bibfnamefont {G.~J.}\ \bibnamefont {Kramer}}, \bibinfo {author}
  {\bibfnamefont {R.}~\bibnamefont {Nazikian}}, \bibinfo {author}
  {\bibfnamefont {D.~C.}\ \bibnamefont {Pace}}, \bibinfo {author}
  {\bibfnamefont {M.}~\bibnamefont {Podest{\`{a}}}}, \bibinfo {author}
  {\bibfnamefont {B.~J.}\ \bibnamefont {Tobias}}, \ and\ \bibinfo {author}
  {\bibfnamefont {M.~A.}\ \bibnamefont {Van~Zeeland}},\ }\href
  {http://stacks.iop.org/0029-5515/57/i=5/a=054001} {\bibfield  {journal}
  {\bibinfo  {journal} {Nucl. Fusion}\ }\textbf {\bibinfo {volume} {57}},\
  \bibinfo {pages} {054001} (\bibinfo {year} {2017})}\BibitemShut {NoStop}%
\bibitem [{\citenamefont {Berk}\ \emph {et~al.}(1995)\citenamefont {Berk},
  \citenamefont {Breizman}, \citenamefont {Fitzpatrick},\ and\ \citenamefont
  {Wong}}]{Berk1995LBQ}%
  \BibitemOpen
  \bibfield  {author} {\bibinfo {author} {\bibfnamefont {H.}~\bibnamefont
  {Berk}}, \bibinfo {author} {\bibfnamefont {B.}~\bibnamefont {Breizman}},
  \bibinfo {author} {\bibfnamefont {J.}~\bibnamefont {Fitzpatrick}}, \ and\
  \bibinfo {author} {\bibfnamefont {H.}~\bibnamefont {Wong}},\ }\href
  {http://stacks.iop.org/0029-5515/35/i=12/a=I30} {\bibfield  {journal}
  {\bibinfo  {journal} {Nucl. Fusion}\ }\textbf {\bibinfo {volume} {35}},\
  \bibinfo {pages} {1661} (\bibinfo {year} {1995})}\BibitemShut {NoStop}%
\bibitem [{\citenamefont {Hickernell}(1984)}]{Hickernell1984}%
  \BibitemOpen
  \bibfield  {author} {\bibinfo {author} {\bibfnamefont {F.~J.}\ \bibnamefont
  {Hickernell}},\ }\href {\doibase 10.1017/S0022112084001178} {\bibfield
  {journal} {\bibinfo  {journal} {J. Fluid Mech.}\ }\textbf {\bibinfo {volume}
  {142}},\ \bibinfo {pages} {431} (\bibinfo {year} {1984})}\BibitemShut
  {NoStop}%
\bibitem [{\citenamefont {Gorelenkov}\ \emph {et~al.}(2019)\citenamefont
  {Gorelenkov}, \citenamefont {Duarte}, \citenamefont {Collins}, \citenamefont
  {Podest\`{a}},\ and\ \citenamefont {White}}]{GorelenkovPoP2019}%
  \BibitemOpen
  \bibfield  {author} {\bibinfo {author} {\bibfnamefont {N.~N.}\ \bibnamefont
  {Gorelenkov}}, \bibinfo {author} {\bibfnamefont {V.~N.}\ \bibnamefont
  {Duarte}}, \bibinfo {author} {\bibfnamefont {C.}~\bibnamefont {Collins}},
  \bibinfo {author} {\bibnamefont {Podest\`{a}}}, \ and\ \bibinfo {author}
  {\bibfnamefont {R.~B.}\ \bibnamefont {White}},\ }\href@noop {} {\bibfield
  {journal} {\bibinfo  {journal} {Phys. Plasmas}\ ,\ \bibinfo {pages}
  {(accepted)}} (\bibinfo {year} {2019})}\BibitemShut {NoStop}%
\bibitem [{\citenamefont {Berk}\ \emph {et~al.}(1997)\citenamefont {Berk},
  \citenamefont {Breizman},\ and\ \citenamefont {Pekker}}]{BerkPPR1997}%
  \BibitemOpen
  \bibfield  {author} {\bibinfo {author} {\bibfnamefont {H.~L.}\ \bibnamefont
  {Berk}}, \bibinfo {author} {\bibfnamefont {B.~N.}\ \bibnamefont {Breizman}},
  \ and\ \bibinfo {author} {\bibfnamefont {M.}~\bibnamefont {Pekker}},\
  }\href@noop {} {\bibfield  {journal} {\bibinfo  {journal} {Plasma Phys.
  Rep.}\ }\textbf {\bibinfo {volume} {23}},\ \bibinfo {pages} {778} (\bibinfo
  {year} {1997})}\BibitemShut {NoStop}%
\bibitem [{\citenamefont {Bhatnagar}\ \emph {et~al.}(1954)\citenamefont
  {Bhatnagar}, \citenamefont {Gross},\ and\ \citenamefont {Krook}}]{Krook1954}%
  \BibitemOpen
  \bibfield  {author} {\bibinfo {author} {\bibfnamefont {P.~L.}\ \bibnamefont
  {Bhatnagar}}, \bibinfo {author} {\bibfnamefont {E.~P.}\ \bibnamefont
  {Gross}}, \ and\ \bibinfo {author} {\bibfnamefont {M.}~\bibnamefont
  {Krook}},\ }\href {https://link.aps.org/doi/10.1103/PhysRev.94.511}
  {\bibfield  {journal} {\bibinfo  {journal} {Phys. Rev.}\ }\textbf {\bibinfo
  {volume} {94}},\ \bibinfo {pages} {511} (\bibinfo {year} {1954})}\BibitemShut
  {NoStop}%
\bibitem [{\citenamefont {Trubnikov}(1965)}]{trubnikov1965particle}%
  \BibitemOpen
  \bibfield  {author} {\bibinfo {author} {\bibfnamefont {B.~A.}\ \bibnamefont
  {Trubnikov}},\ }\href@noop {} {\bibfield  {journal} {\bibinfo  {journal}
  {Rev. Plasma Phys.}\ }\textbf {\bibinfo {volume} {1}},\ \bibinfo {pages}
  {105} (\bibinfo {year} {1965})}\BibitemShut {NoStop}%
\bibitem [{\citenamefont {Sanz-Orozco}\ \emph {et~al.}(2018)\citenamefont
  {Sanz-Orozco}, \citenamefont {Berk}, \citenamefont {Faganello}, \citenamefont
  {Idouakass},\ and\ \citenamefont {Wang}}]{SanzNF2018}%
  \BibitemOpen
  \bibfield  {author} {\bibinfo {author} {\bibfnamefont {D.}~\bibnamefont
  {Sanz-Orozco}}, \bibinfo {author} {\bibfnamefont {H.}~\bibnamefont {Berk}},
  \bibinfo {author} {\bibfnamefont {M.}~\bibnamefont {Faganello}}, \bibinfo
  {author} {\bibfnamefont {M.}~\bibnamefont {Idouakass}}, \ and\ \bibinfo
  {author} {\bibfnamefont {G.}~\bibnamefont {Wang}},\ }\href
  {http://stacks.iop.org/0029-5515/58/i=8/a=082012} {\bibfield  {journal}
  {\bibinfo  {journal} {Nuclear Fusion}\ }\textbf {\bibinfo {volume} {58}},\
  \bibinfo {pages} {082012} (\bibinfo {year} {2018})}\BibitemShut {NoStop}%
\bibitem [{\citenamefont {Duarte}\ and\ \citenamefont
  {Gorelenkov}(2019)}]{duarte2018analytical}%
  \BibitemOpen
  \bibfield  {author} {\bibinfo {author} {\bibfnamefont {V.~N.}\ \bibnamefont
  {Duarte}}\ and\ \bibinfo {author} {\bibfnamefont {N.~N.}\ \bibnamefont
  {Gorelenkov}},\ }\href {\doibase 10.1088/1741-4326/ab0135} {\bibfield
  {journal} {\bibinfo  {journal} {Nucl. Fusion}\ }\textbf {\bibinfo {volume}
  {59}},\ \bibinfo {pages} {044003} (\bibinfo {year} {2019})}\BibitemShut
  {NoStop}%
\bibitem [{\citenamefont {Van~Vleck}\ and\ \citenamefont
  {Weisskopf}(1945)}]{vanVleck1945shape}%
  \BibitemOpen
  \bibfield  {author} {\bibinfo {author} {\bibfnamefont {J.~H.}\ \bibnamefont
  {Van~Vleck}}\ and\ \bibinfo {author} {\bibfnamefont {V.~F.}\ \bibnamefont
  {Weisskopf}},\ }\href
  {https://journals.aps.org/rmp/abstract/10.1103/RevModPhys.17.227} {\bibfield
  {journal} {\bibinfo  {journal} {Rev. Mod. Phys.}\ }\textbf {\bibinfo {volume}
  {17}},\ \bibinfo {pages} {227} (\bibinfo {year} {1945})}\BibitemShut
  {NoStop}%
\bibitem [{\citenamefont {Peach}(1981)}]{peach1981theory}%
  \BibitemOpen
  \bibfield  {author} {\bibinfo {author} {\bibfnamefont {G.}~\bibnamefont
  {Peach}},\ }\href
  {https://www.tandfonline.com/doi/abs/10.1080/00018738100101467} {\bibfield
  {journal} {\bibinfo  {journal} {Adv. Phys.}\ }\textbf {\bibinfo {volume}
  {30}},\ \bibinfo {pages} {367} (\bibinfo {year} {1981})}\BibitemShut
  {NoStop}%
\bibitem [{\citenamefont {Dupree}(1966)}]{Dupree1966}%
  \BibitemOpen
  \bibfield  {author} {\bibinfo {author} {\bibfnamefont {T.~H.}\ \bibnamefont
  {Dupree}},\ }\href {\doibase 10.1063/1.1761932} {\bibfield  {journal}
  {\bibinfo  {journal} {Phys. Fluids}\ }\textbf {\bibinfo {volume} {9}},\
  \bibinfo {pages} {1773} (\bibinfo {year} {1966})}\BibitemShut {NoStop}%
\bibitem [{\citenamefont {Krommes}(2002)}]{krommes2002fundamental}%
  \BibitemOpen
  \bibfield  {author} {\bibinfo {author} {\bibfnamefont {J.~A.}\ \bibnamefont
  {Krommes}},\ }\href
  {https://www.sciencedirect.com/science/article/pii/S0370157301000667}
  {\bibfield  {journal} {\bibinfo  {journal} {Phys. Rep.}\ }\textbf {\bibinfo
  {volume} {360}},\ \bibinfo {pages} {1} (\bibinfo {year} {2002})}\BibitemShut
  {NoStop}%
\bibitem [{\citenamefont {White}\ \emph {et~al.}(2019)\citenamefont {White},
  \citenamefont {Duarte}, \citenamefont {Gorelenkov},\ and\ \citenamefont
  {Meng}}]{WhiteDuarteGorMengPoP2019}%
  \BibitemOpen
  \bibfield  {author} {\bibinfo {author} {\bibfnamefont {R.~B.}\ \bibnamefont
  {White}}, \bibinfo {author} {\bibfnamefont {V.~N.}\ \bibnamefont {Duarte}},
  \bibinfo {author} {\bibfnamefont {N.~N.}\ \bibnamefont {Gorelenkov}}, \ and\
  \bibinfo {author} {\bibfnamefont {G.}~\bibnamefont {Meng}},\ }\href {\doibase
  10.1063/1.5088598} {\bibfield  {journal} {\bibinfo  {journal} {Phys.
  Plasmas}\ }\textbf {\bibinfo {volume} {26}},\ \bibinfo {pages} {032508}
  (\bibinfo {year} {2019})}\BibitemShut {NoStop}%
\end{thebibliography}
\addcontentsline{toc}{section}{\refname}

\end{document}